\documentclass[prd,twocolumn,showpacs,superscriptaddress,preprintnumbers,amsmath,amssymb,nofootinbib]{revtex4}

\usepackage{graphicx}
\usepackage{dcolumn}
\usepackage{bm}
\usepackage{color}
\usepackage{epsfig}
\usepackage{amssymb}
\usepackage{amsmath}

\usepackage[usenames, dvipsnames]{xcolor}
\usepackage{url}
\usepackage[linktocpage=true]{hyperref}
\hypersetup{
   colorlinks   = true,
    linkcolor= Blue,
    citecolor= Blue,
    urlcolor= Blue}

\newcommand{\MS}{\ensuremath{\overline{\text{MS}}}}
\newcommand{\bea}{\begin{eqnarray}}
\newcommand{\eea}{\end{eqnarray}}
\newcommand{\nn}{\nonumber}

\newcommand{\eq}[1]{Eq.~(\ref{#1})}

\newcommand{\be}{\begin{equation}}
\newcommand{\ee}{\end{equation}}

\begin{document}

\title{\boldmath \Large\bf (${\cal O}_8$,\,${\cal O}_8$) contribution to $\bar{B} \to X_s\gamma \gamma$ at $O(\alpha_s)$}

\author{H. M. Asatrian}
\email[Electronic address:~]{hrachia@itp.unibe.ch}
\affiliation{Yerevan Physics Institute, 0036 Yerevan, Armenia}
\author{C. Greub}
\email[Electronic address:~]{greub@itp.unibe.ch}
\affiliation{Albert Einstein Center for Fundamental Physics, Institute
  for Theoretical Physics,\\ University of Bern, CH-3012 Bern,
  Switzerland}
\author{A. Kokulu}
\email[Electronic address:~]{akokulu@liverpool.ac.uk}
 \affiliation{Department of Mathematical Sciences, University of Liverpool,\\ L69 3BX Liverpool, United Kingdom} 
 %



\begin{abstract}

In this analysis, we present the contribution associated with the
chromomagnetic dipole operator ${\cal O}_8$
to the double differential decay width $d\Gamma/(ds_1 ds_2)$ for the inclusive
process $\bar{B} \to X_s \gamma \gamma$.  
The kinematical variables $s_1$ and $s_2$ are defined as
$s_i=(p_b - q_i)^2/m_b^2$, where $p_b$, $q_1$, $q_2$ are the momenta
of $b$-quark and two photons. This contribution (taken at tree level) is of
order $\alpha_s$, like the recently calculated QCD corrections to the
contribution of the operator ${\cal O}_7$.
In order to regulate possible collinear singularities of one of the photons
with the strange quark, 
we introduce a nonzero mass $m_{s}$ for the strange quark. 
Our results are obtained for exact $m_{s}$, which we interpret as
a constituent mass being varied between 400 and 600 $\rm MeV$. 
Numerically it turns out that the effect of the (${\cal O}_8,{\cal O}_8$) contribution to the branching ratio of $\bar{B} \to X_s
\gamma \gamma$ does not exceed $+0.1\, \%$ for any kinematically allowed value of our physical cutoff parameter $c$,
confirming the expected suppression of this contribution relative to the QCD
corrections to $d\Gamma_{77}/(ds_1 \, ds_2)$ 
\cite{Asatrian:2011ta,Asatrian:2014mwa}.

\end{abstract}


\preprint{LTH 1061}

\maketitle


\section{Introduction}
\label{sec:Introduction}

Inclusive rare $B$-meson decays are known to be a unique source of indirect information about 
physics at scales of several hundred GeV. In the standard model (SM) all these processes 
proceed through loop diagrams and thus are relatively suppressed. In the extensions 
of the SM the contributions stemming from the diagrams with ``new'' 
particles in the loops can be comparable or even larger than the contribution from 
the SM. Thus getting experimental information on rare decays puts strong 
constraints on the extensions of the SM or can even lead to a  
disagreement with the SM predictions, providing evidence for some "new physics''. 

To make a rigorous comparison between experiment and theory, precise
SM calculations for the (differential) decay rates are mandatory. 
While the
branching ratios for $\bar{B} \to X_s \gamma$ \cite{Misiak:2006zs}
and $\bar{B} \to X_s \ell^+
\ell^-$ are known today even to
next-to-next-to-leading logarithmic (NNLL) precision (for reviews, see
\cite{Hurth:2010tk,Buras:2011we} and \cite{Misiak:2015xwa} for recent updated predictions on radiative decay modes of $B$ meson),
other branching ratios, like the one for $\bar{B} \to X_s \gamma
\gamma$ discussed in this paper, were only known to leading logarithmic
(LL) precision in the SM
\cite{Simma:1990nr,Reina:1996up,Reina:1997my,Cao:2001uj}. 
As the process $\bar{B} \to X_s \gamma \gamma$ is expected to be measured at the
planned Super $B$-factory in Japan (SuperKEKB)
\cite{Aushev:2010bq,superKEKB:online}, we recently completed 
first steps towards a next-to-leading logarithmic (NLL) result for this decay
\cite{Asatrian:2011ta,Asatrian:2014mwa}, by working out QCD corrections to the numerically
important (${\cal O}_7$,~${\cal O}_7$) contribution. 

In this paper, we go one step further and provide the self-interference
contribution to $\bar{B} \to X_s \gamma \gamma$ stemming from the chromomagnetic dipole operator ${\cal O}_8$
which starts at order $\alpha_s$. Although a naive estimate suggests that this
contribution is suppressed by a factor of  $|C_{8}^{{\rm eff}}
Q_{d}/C_{7}^{{\rm eff}}|^{2} \sim 1/36$ relative to the QCD corrections to the (${\cal O}_7$,~${\cal O}_7$)
interference, a more detailed investigation is in order: In both cases 
(${\cal O}_7$ and ${\cal O}_8$), one of the two photons can be emitted from
the strange quark in a collinear way, leading to contributions involving
$\log(m_{s}/m_{b})$ terms.\footnote{We interpret $m_s$ to be a constituent
mass, varying it between 400 and 600 MeV.} Concerning the other photon, the
two cases differ, however. Unlike in the ${\cal O}_7$, the second photon can also be emitted
from the $s$-quark in the ${\cal O}_8$ case. While a fully collinear emission
of both photons is excluded by our cuts (see later), a leftover enhancement effect could
still apply in the ${\cal O}_8$ case and thereby milder the naive suppression
factor. As the average energies of the two photons are not very high, there
might be a second effect related to the different infrared structure
($1/E_{\gamma}$-terms) of the two cases, which also potentially milders the naive suppression
factor given above. We feel that these considerations motivate a detailed
evaluation of the $({\cal O}_8,{\cal O}_8)$-interference contribution.

The starting point of our calculation is the effective Hamiltonian,
obtained by integrating out the heavy particles in the SM, leading to
\be
 {\cal H}_{eff} = - \frac{4 G_F}{\sqrt{2}} \,V_{ts}^\star V_{tb} 
   \sum_{i=1}^8 C_i(\mu) {\cal O}_i(\mu)  \, ,
\label{Heff}
\ee
where we use the operator basis introduced in \cite{Chetyrkin:1996vx}:
\vspace{0.2cm}
\be
\begin{array}{llll}
{\cal O}_1 \, &\!= 
 (\bar{s}_L \gamma_\mu T^a c_L)\, 
 (\bar{c}_L \gamma^\mu T_a b_L)\,, 
              \,   \quad  \quad \quad   \\[1.002ex]
{\cal O}_2 \, &= 
 (\bar{s}_L \gamma_\mu c_L)\, 
 (\bar{c}_L \gamma^\mu b_L)\,,   \\[1.002ex]
{\cal O}_3 \, &\! = 
 (\bar{s}_L \gamma_\mu b_L) 
 \sum_q
 (\bar{q} \gamma^\mu q)\,, 
                 \quad  \quad \quad \quad \quad   \\[1.002ex]
{\cal O}_4 \, &= 
 (\bar{s}_L \gamma_\mu T^a b_L) 
 \sum_q
 (\bar{q} \gamma^\mu T_a q)\,,  \\[1.002ex]
{\cal O}_5 \, &\! = 
 (\bar{s}_L \gamma_\mu \gamma_\nu \gamma_\rho b_L) 
 \sum_q
 (\bar{q} \gamma^\mu \gamma^\nu \gamma^\rho q)\,, 
                 \quad \,   \\[1.002ex]
{\cal O}_6 \, &=
 (\bar{s}_L \gamma_\mu \gamma_\nu \gamma_\rho T^a b_L) 
 \sum_q
 (\bar{q} \gamma^\mu \gamma^\nu \gamma^\rho T_a q)\,,  \\[1.002ex]
{\cal O}_7 \, &\! = 
  \frac{e}{16\pi^2} \,\left[     
 \bar{s} \sigma^{\mu\nu}   \left( \bar{m}_b \,R\,   +  \, \bar{m}_s \,L  \right)  F_{\mu\nu} b  \right]  \,,   \\[1.002ex]
{\cal O}_8 \, &\! = 
  \frac{g_{s}}{16\pi^2} \,\left[     
 \bar{s} \sigma^{\mu\nu}   \left( \bar{m}_b \,R\,   +  \, \bar{m}_s \,L  \right)  T^a G^a_{\mu\nu} b  \right]  \, .
\end{array} 
\label{opbasis}
\vspace{0.2cm}
\ee

The symbols $T^a$ ($a=1,8$) denote the $SU(3)$ color generators; 
$g_s$ and $e$ denote the strong and electromagnetic coupling constants.
In \eq{opbasis}, 
$\bar{m}_b$ and $\bar{m}_s$ are the running $b$ and $s$-quark masses 
in the $\MS$-scheme at the renormalization scale $\mu$. 
We keep the exact dependence on the strange-quark mass in our calculation.
Further, as we are not interested in {\it CP}-violation effects in the present paper, we 
exploited the unitarity of the Cabibbo--Kobayashi--Maskawa (CKM) matrix and
neglected $V_{ub} V_{us}^*$ (as $V_{ub} V_{us}^* \ll V_{tb} V_{ts}^* $) when writing
\eq{Heff}.

While the Wilson coefficients $C_i(\mu)$ appearing in \eq{Heff}
have been known to sufficient precision at the low scale $\mu \sim m_b$
for a long time (see e.g. the reviews \cite{Hurth:2010tk,Buras:2011we}
and references therein), the matrix elements 
$\langle s \gamma \gamma|{\cal  O}_i|b\rangle$ and 
$\langle s \gamma \gamma \, g|{\cal  O}_i|b\rangle$, 
which in a NLL calculation are needed to order
$g_s^2$ and $g_s$, respectively, are only partially known now (see
\cite{Asatrian:2011ta,Asatrian:2014mwa} for the details of the provided
contributions 
and \cite{Kokulu:2014qqa} for a recent summary). Calculating the
$({\cal O}_i,{\cal O}_j)$-interference contributions for the
differential distributions at order
$\alpha_s$ is in many respects of similar complexity as the
calculation of the photon energy spectrum in $\bar{B} \to X_s \gamma$ 
at order $\alpha_s^2$
needed for the NNLL computation. There, the individual
interference contributions, which all involve extensive calculations, were
published in separate papers, sometimes even by two independent groups
(see e.g. \cite{Melnikov:2005bx,Asatrian:2006sm}).
It therefore cannot be expected that the NLL results for the
differential distributions related to $\bar{B} \to X_s \gamma \gamma$ are 
given in a single paper. As a next step in the NLL enterprise, we
derive in the present paper the $({\cal O}_8,{\cal O}_8)$-interference
contribution (which starts at order $\alpha_s$) to the double 
differential decay width $d\Gamma/(ds_1 ds_2)$. 
The variables $s_1$
and $s_2$ are defined as $s_i=(p_b-q_i)^2/m_b^2$, where $p_b$
and $q_i$ denote the four-momenta of the $b$-quark and the two
photons, 
respectively.

At order $\alpha_s$ there are only contributions to $d\Gamma_{88}/(ds_1 ds_2)$
with four particles ($s$-quark, two photons and a gluon) in the final state.
These contributions correspond to specific cuts of the $b$-quark
self energy at order $\alpha^2 \times \alpha_s$, involving twice the
operator ${\cal O}_8$. As there are additional cuts, which contain for
example only one photon, our observable cannot be obtained using the
optical theorem, i.e., by taking the absorptive part of the $b$-quark
self energy at three loops. We therefore calculate the mentioned 
contributions with four particles in the final state individually.

When calculating the contribution of ${\cal O}_8$ to $d\Gamma/(ds_1 ds_2)$,
we restrict ourselves (as in refs. \cite{Asatrian:2011ta,Asatrian:2014mwa}) to the region in the $(s_1,s_2)$-plane
which is also accessible to three body decays $b \to s \gamma \gamma$
(associated e.g. with the tree-level contribution of ${\cal O}_7$), i.e.,
\be
s_1 > x_4 \, ; \, 
s_2 > x_4 \, ; \,
s_1 + s_2 < 1+ x_4 \, ; \, 
s_1 s_2 > x_4 \, ,
\label{eq:cutsA}
\ee
where $x_4=(m_s/m_b)^2$.
The energies $E_1$ and $E_2$ in the rest frame of the $b$-quark of the two photons are 
related to $s_1$ and $s_2$ in a simple way: $s_i=1-2 \, E_i/m_b$.
As the energies $E_i$ of the photons have to be away from zero in order to
be observed, the values of $s_1$ and $s_2$ should be considered to be
smaller than one. Furthermore, in order to see two separate photons, their invariant mass should
also be away from zero. All these requirements can be implemented in terms of one 
physical cut parameter $c$ ($c>0$), by demanding\footnote{The normalized
  invariant mass squared $s=(q_1+q_2)/m_b^2$ of the two photons can be written
  as $s=1-s_1-s_2+s_3$, where $s_3$ is the normalized hadronic mass squared.} 
\bea
s_1\geq c~ ,\,\, \,  s_2 \geq c~,\, \,\, 1-s_{1}-s_{2} \geq c   \,.
\label{eq:cuts}
\eea
The kinematical region in the $(s_1,s_2)$-plane, which we take into account in this
paper, therefore corresponds to the intersection of the regions given in eqs.
(\ref{eq:cutsA}) and (\ref{eq:cuts}). For explicit formulas representing this
intersection, we refer to the appendix.

Imposing these cuts, the photons do not become soft in our case, while one of
them can become collinear with the strange quark. 
This implies that in the final result a single logarithm of $m_{s}$ survives.
The only source for such $\log(m_s)$ terms in our result is the mentioned
 collinear emission of the photons from the $s$ quark. In particular, we emphasize that the
 $({\cal O}_8,{\cal O}_8)$-contribution to the double differential decay width
 does not become singular when the gluon and the strange quark become
 collinear, since the gluon is
 emitted from the effective operator ${\cal O}_8$ directly and therefore there
 is no propagator denominator of the form $(p_s + p_g)^2$ which could become singular. In addition,
 soft-gluon related singularities also do not appear in this case (the matrix
 element associated with ${\cal O}_8$ even goes to zero when the gluon energy
 tends to $0$). The absence of singularities generated by soft and/or
 collinear gluons is related to the fact that concerning QCD our
 observable (i.e. the triple or double differential decay width), based on the
 full effective Hamiltonian, is fully inclusive and therefore nonsingular.
 We also stressed this fact in \cite{Asatrian:2014mwa}, where
 the  $({\cal O}_7,{\cal O}_7)$-contribution was worked out. In this case
 there were gluon induced singularities in the virtual and bremsstrahlung
 corrections, but they canceled when combined 
 as a consequence of the Kinoshita-Lee-Nauenberg (KLN) theorem. This means that the origin of
 $\log(m_s)$ terms is from collinear photon emission only. Note that concerning QED our
 observable is not fully inclusive, because we want to observe exactly two
 photons in the final state; therefore $\log(m_s)$ terms remain.
 A further remark on the numerical $m_s$-dependence is in order: The  $({\cal O}_7,{\cal O}_7)$-contribution
 to the double differential decay width starts at order $\alpha_s^0$. This
 leading contribution does not contain $\log(m_s)$ terms when applying the
 kinematical cuts discussed above. Only at order $\alpha_s^1$ 
 terms $\sim \log(m_s)$ appear, because one of the photons can become collinear with the
 strange quark. As a consequence, we expect the relative $m_s$-dependence of the
 $({\cal O}_7,{\cal O}_7)$ contribution to be smaller than the corresponding dependence of the
 $({\cal O}_8,{\cal O}_8)$ contribution, because the latter only starts at order
 $\alpha_s^1$. In other words the $m_s$-dependence of the complete double
 differential decay width will be smaller than the one which is only based on
 the  $({\cal O}_8,{\cal O}_8)$ contribution discussed in this paper.

The main goal of this paper is to work out $d\Gamma_{88}/(ds_1 ds_2)$ as a further  
ingredient towards a systematic NLL prediction for the decay rate of $\bar{B} \to
X_s \gamma \gamma$. 
For similar analysis for the case of $\bar{B} \to X_s \gamma$, one can see
e.g. \cite{Ferroglia:2010xe,Misiak:2010tk,Benzke:2010js,Kapustin:1995fk,Ali:1995bi}.
 
{In this regard, we employ in our calculation a finite
  strange-quark mass $m_s$ which we interpret to be of
  constituent type in the numerics. This approach
  has also been adopted previously, e.g. by Kaminski {\it et al.} in
  \cite{Kaminski:2012eb} and Asatrian and Greub in \cite{Asatrian:2013raa,Asatrian:2014mwa}. The experience gained in these references shows that
  the constituent mass approach gives
  results which are similar to those when using fragmentation functions
  \cite{Asatrian:2013raa}. Therefore, we believe that this method is sufficient
  to obtain an estimate of the $({\cal O}_8,{\cal O}_8)$-interference
  contribution. While the fragmentation approach seems better from the
  theoretical point of view, it is not clear that it leads to
  better final results in practice, because  the fragmentation
  functions (for $s\to\gamma$ or $g\to\gamma$) suffer from experimental
  uncertainties, as pointed out in \cite{Asatrian:2013raa}. An alternative could be to look at the version with
  ``isolated photons'' a la Frixione \cite{Frixione:1998jh} which corresponds,
  however, to a slightly different observable. Such an approach is beyond the
  scope of the present paper and is left for future studies.}

Before moving to the detailed organization of our paper, we should
mention that the inclusive double radiative process $\bar{B} \to X_s \gamma
\gamma$ has also been explored in several extensions of the SM
\cite{Reina:1996up,Gemintern:2004bw,Cao:2001uj}. Also 
the corresponding exclusive modes, $B_s \to \gamma \gamma$  
and $B\to K \gamma \gamma$, have been 
examined before, both in the SM
\cite{Reina:1997my,Chang:1997fs,Hiller:1997ie,Bosch:2002bv,Bosch:2002bw,
Hiller:2004wc,Hiller:2005ga,Lin:1989vj,Herrlich:1991bq,Choudhury:2002yu} and
in its extensions
\cite{Gemintern:2004bw,Hiller:2004wc,Hiller:2005ga,Aliev:1997uz,Bertolini:1998hp,Bigi:2006vc,Devidze:1998hy,Aliev:1993ea,Xiao:2003jn,Huo:2003cj,Chen:2011te,Qin:2009zzb}. 
We should add that the long-distance resonant effects were
also discussed in the literature (see e.g. \cite{Reina:1997my} and the
references therein). 
Finally, the effects of photon emission from the spectator quark in
the $B$-meson were discussed in \cite{Chang:1997fs,Hiller:2004wc,Ignatiev:2003qm}. 

The remainder of this paper is organized as follows. In section \ref{sec:bremsstrahlung}
the calculation of the $({\cal O}_8,{\cal O}_8)$-contribution to the double
differential decay width $d\Gamma/(ds_1 ds_2)$ is presented.  
To regulate the configurations where photons are emitted from the $s$ quark in a
collinear way, a finite strange-quark mass $m_{s}$ is introduced.
This way the collinear singularities manifest themselves as $\log(m_{s})$
terms in our final result, which reflects the feature for the photons having hadronic substructure. 
In section \ref{sec:numerics} we illustrate the numerical impact of the
$({\cal O}_8,{\cal O}_8)$-contribution to the double differential width and the
total decay width (depending on a kinematical cut).
The main text of our paper ends with a short summary in section
\ref{sec:summary}. In the appendix \ref{sec:appendix}, we give the explicit formulas defining the four-particle phase-space region considered in this paper together with the explicit
expressions for the master integrals (MIs) appearing in our calculation.

\section{$({\cal O}_8,{\cal O}_8)$ contribution to the double differential spectrum
  $d\Gamma/(ds_1 ds_2)$ at ${\cal O}(\alpha_{s})$}
\label{sec:bremsstrahlung}

\begin{figure}[h]
\begin{center}
\includegraphics[width=0.4\textwidth]{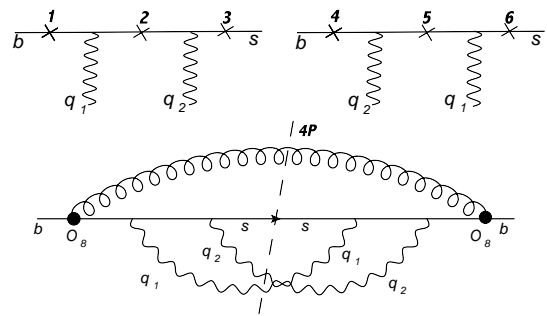}
\caption{On the first line the diagrams defining the
${\cal O}_8$ contribution to $b \to s g \gamma \gamma$ are shown
at the amplitude level. The crosses in the graphs stand
for the possible emission places of the gluon (emerging from the operator ${\cal O}_8$).
On the second line the contribution to the decay width corresponding
to the interference of  diagram 1 with diagram 4 is illustrated. This sample
interference diagram gives rise to 
$\log\left( m_{s}/m_{b}\right)$ terms due to collinear configurations of one of the photons with the $s$ quark.}
\label{fig:amplitudebrems}
\end{center}
\end{figure}
We now turn to the calculation of the ${\cal O}_8$ self-interference contribution to the decay width for $\bar{B} \to X_s \gamma
\gamma$, which is based on the partonic process $b \to s g \gamma
\gamma$, where $g$ denotes a gluon. Although this is only a tree-level computation at order $\alpha_s$,
it is quite complicated because of the four particles in the final state, one
of them being massive (the strange quark). 

Before going into detail, we mention that the kinematical range of
the variables $s_1=(p_b-q_1)^2/m_b^2$ and $s_2=(p_b-q_2)^2/m_b^2$ is larger in
the $1 \to 4$ process considered in this section than the range given in \eq{eq:cutsA}, which corresponds
to the $1 \to 3$ process $b \to s \gamma \gamma$.
Nevertheless, we restrict ourselves to the range which corresponds to the
intersection of the regions given in \eq{eq:cutsA} and \eq{eq:cuts}, as
we also did in \cite{Asatrian:2011ta,Asatrian:2014mwa} when considering
virtual and bremsstrahlung corrections to the ${\cal O}_7$-contribution. For explicit formulas of the considered $(s_1,s_2)$-region, we refer to 
\eq{eq:PScases} in the appendix.

The diagrams defining the ${\cal O}_8$ contribution at the
amplitude level are shown in the first line of
Fig. \ref{fig:amplitudebrems}. The amplitude squared, needed to get the (double differential) decay width, can be written
as a sum of interferences of the different diagrams shown on the first line in
Fig. \ref{fig:amplitudebrems}. One such interference is shown on the
second line of the same figure. The four-particle final state is
described by five independent kinematical variables; $s_1$ and $s_2$ are just
two of them.

In the present paper, we worked out in a first step the triple differential
spectrum $d\Gamma_{88}/(ds_1 \, ds_2\, ds_{3})$, where
$s_{3}=(p_{s}+p_{g})^{2}/m_{b}^{2}$ is the normalized hadronic mass squared
and $p_{g}$ is the final state gluon momentum. At this level, we computed
the resulting MIs numerically for exact $m_{s}$ (see
section \ref{sec:appendix} for their explicit expressions). To get the double
differential spectrum $d\Gamma_{88}/(ds_1 \, ds_2)$ we then integrated over
$s_{3}$ in its range $s_{3}\, \in \, \left[  m_{s}^{2}/m_{b}^{2},\,
  s_{1}.s_{2}\right]$. 

Last, as the various steps of the calculation are similar to those in
Ref.~\cite{Asatrian:2011ta}, we refer to section $7$ of that paper for more
details on the techniques applied. Also, we refer to appendix B of
Ref.~\cite{Asatrian:2014mwa} for a useful
parametrization of the four-particle phase-space for the case where one of the
particles is massive, which is based on the work in
Ref.~\cite{Asatrian:2006ph}.

\begin{table}[htbp]
\begin{minipage}{3in}
\centering 
  \begin{tabular}{@{}|c|c|c|c@{}|}
 \hline
 Parameter & Value \\   \hline \hline

$\rm BR_{sl}^{exp}$& $0.1049$   \\ \hline

$\rm m_{b}$& $4.8$~GeV   \\ \hline

$\rm m_{c}/m_{b}$ & $0.29$   \\ \hline




$\rm G_{F}$&$1.16637\times10^{-5}$~\text{GeV}$^{-2}$  \\ \hline

$\rm V_{cb} $&$0.04$    \\  \hline

$\rm V_{tb} V_{ts}^* $&$0.04$    \\  \hline

$\rm {\alpha_{\rm (em)}}^{-1}$ & $137$    \\ \hline


\end{tabular}
 \end{minipage} ~~~~~~~~~~~~~
\begin{minipage}{3in}
  \vspace{0.5cm}
  \begin{tabular}{|c|c|c|}
\hline
 & $C_{8,eff}^{0}(\mu)$ ~&~ $\alpha_s(\mu)$ 
\\   \hline \hline

$\mu= M_W$  & ~$-0.09739$ ~&   $0.1213$     \\ \hline 

 $\mu=2 \, m_{b}$  & $-0.13516$~ & ~$0.1818$    \\ \hline 

 $\mu=m_{b}$  & $-0.14905$~ & ~$0.2175$      \\ \hline

 $\mu=m_{b}/2$  &  $-0.16529$~ & ~ $0.2714$           \\ \hline \hline

\end{tabular}  
    \end{minipage}
\caption{ {\bf Upper:} Relevant input parameters used in this paper. {\bf Lower:} The Wilson coefficient $C_{8,eff}(\mu)$ and $\alpha_s(\mu)$ at
   different values of the renormalization scale $\mu$.}
\label{tab12:wilson_input}
\end{table}   
\vspace{1mm}

\section{Numerical illustrations}\label{sec:numerics}
In the previous section we described the calculation for the $({\cal O}_8,{\cal O}_8)$ contribution to the double differential decay
width for $\bar{B} \to X_s \gamma \gamma$ at NLL precision.

The Wilson coefficient $C_{8,eff}(\mu)$ at the low scale\footnote{At NLL
  precision, $C_{8,eff}(\mu)$ is needed only up to order $\alpha_s^{0}$,
  because the square of the matrix element $\langle s g \gamma \gamma| {\cal O}_8|
  b \rangle$ starts at order $\alpha_s^1$. 
 Furthermore, for our current purpose we identify the
 $\overline{\mbox{MS}}$ mass $\bar{m}_b(\mu)$  
with the corresponding pole mass.} 
\be
\nn
C_{8,eff}(\mu) = C_{8,eff}^{0}(\mu_{b})
\ee
has been known for a long time (see Ref.~\cite{Chetyrkin:1996vx} and
references therein).
Numerical values for
the input parameters and for this Wilson coefficient at various values
for the scale $\mu$, together with the
numerical values of $\alpha_s(\mu)$, are given in upper and lower panels of Table~\ref{tab12:wilson_input}, respectively.

To stress that the $({\cal O}_8,{\cal O}_8)$-contribution to $d\Gamma/(ds_1 ds_2)$ only
starts at the NLL level, we write
\be
\frac{d\Gamma_{88}}{ds_1 ds_2} =  \frac{d\Gamma_{88}^{(1)}}{ds_1 ds_2}
\label{eq:total88a}
\ee  
where $d\Gamma_{88}^{(1)}/(ds_1 ds_2)$ has the form
\begin{eqnarray}
 \frac{d\Gamma_{88}^{(1)}}{ds_1 \, ds_2} &&\,=\, 
\frac{  \alpha^2 \, \bar{m}_b^2(\mu) \, m_{b}^3 \, |C_{8,eff}(\mu)|^2 \, G_F^2  \, 
  |V_{tb} V_{ts}^*|^2 \, Q_d^4}{1024 \, \pi^5} \,  \,  \,  \,  \nonumber \\
&& 
\times \,\frac{\alpha_s}{4 \pi} \, C_{F}\, \,\kappa^{(1)}_{88}(s_{1},s_{2},m_{s}/m_{b})  \, .
\label{eq:double88}
\end{eqnarray}
The function $\kappa^{(1)}_{88}(s_{1},s_{2},m_{s}/m_{b})$, which encodes the
dependence on $s_1$, $s_2$ and on $m_s/m_b$, is too lengthy to be displayed
explicitly. We note that we will keep the exact $m_s$ dependence in our
numerics.

\begin{center}
\begin{figure}[h]
\centering{
\includegraphics[width=0.45\textwidth]{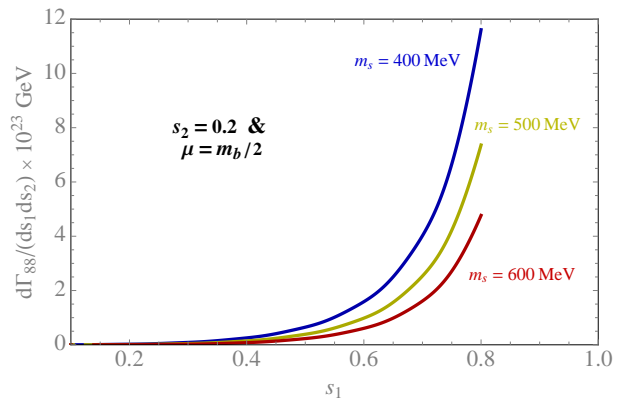}
\caption{$d\Gamma_{88}/(ds_1 \, ds_2)$ [as given in eqs. (\ref{eq:total88a}) and (\ref{eq:double88})] as a
  function of $s_1$ for
  $s_{2}$ fixed at $0.2$, $\mu=m_{b}/2$ and $m_{s}$ varied between
  $400$ and $600$ MeV. The blue(top), yellow(middle) and
  red(bottom) lines show the width when choosing $m_{s}$ to be $400$,
  $500$ and $600$ MeV, respectively.}      
\label{fig:width_o8o8}
}
\end{figure}
\end{center}

In Table~\ref{tab:o8o8BRs}, the impact of $\frac{d\Gamma_{88}^{(1)}}{ds_1 \, ds_2}$ on the branching ratio for $\bar{B} \to X_s \gamma \gamma$ is presented
for various choices of $m_{s}$, $c$ and the scale $\mu$. It is seen that this contribution is much
smaller than the corresponding numbers for the $({\cal O}_7$, ${\cal O}_7)$
contribution (see Table 4 of Ref.~\cite{Asatrian:2014mwa} for comparison).

To obtain the values for the branching ratio in Table~\ref{tab:o8o8BRs} as a function of the cutoff parameter $c$ defined in
\eq{eq:cuts}, we integrate the double differential spectrum over the
corresponding ranges in $s_{1}$ and $s_{2}$ ${\left[\rm see~\eq{eq:PScases}\right]}$, divide by
the semileptonic decay width and multiply with the measured semileptonic
branching ratio. For illustrative purposes, it is sufficient to take the lowest order
formula for the semileptonic decay width [see e.g. Eq.~(6.2) in Ref.~\cite{Asatrian:2014mwa}].

In Fig.~\ref{fig:width_o8o8} we plot $d\Gamma_{88}/(ds_1 \, ds_2)$, calculated
in this paper, as a function of $s_1$, while $s_2$ is kept fixed at
$s_{2}=0.2$. The renormalization scale is chosen to be $\mu=m_{b}/2$ and
$m_{s}$ is varied between $400$ and $600$ MeV. This figure shows that 
$d\Gamma_{88}/(ds_1 \, ds_2)$ is orders of magnitude smaller in size
than $d\Gamma_{77}/(ds_1 \, ds_2)$ (for comparison see Fig. 7
of Ref.~\cite{Asatrian:2014mwa} which is an extended analysis of the work in
Ref.~\cite{Asatrian:2011ta}). For other choices of the scale $\mu$, the
behavior of the spectrum is similar, but even smaller in size.

\begin{figure}[h]
\begin{center}
\includegraphics[width=0.45\textwidth]{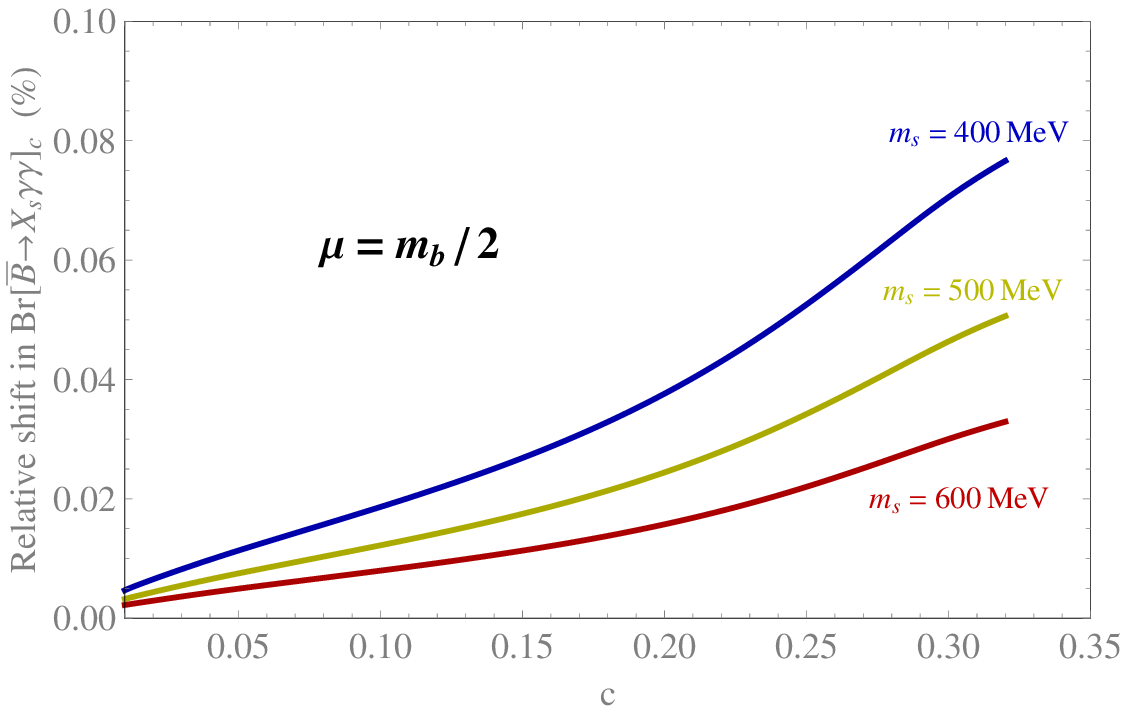}     
\caption{Relative shift ($  {\rm  \frac{   Br[\bar{B}\to X_s\gamma \gamma
      ]_{c}^{88} }{ Br[\bar{B}\to X_s\gamma \gamma ] _{c}} } $) of 
the branching ratio for ${\rm \bar{B}\to X_s\gamma \gamma }$ (in percent)
due to the (${\cal O}_8$\,,\,${\cal O}_8$) contribution as a function
of the cut parameter $c$ for $\mu=m_{b}/2$. The blue(top), yellow(middle)
and red(bottom) lines show the relative shifts when setting $m_{s}=400$ MeV,
$500$ MeV and $600$ MeV, respectively. For other choices of the scale $\mu$
the relative change is even smaller.}      
\label{fig:BR_relativeshift}
\end{center} 
\end{figure}

In Fig.~\ref{fig:BR_relativeshift} we investigate the numerical impact of the 
(${\cal O}_8$\,,\,${\cal O}_8$) contribution on the branching ratio
of $\bar{B}\to X_s\gamma \gamma$ (see the discussion in the
third paragraph of the introduction). More precisely,
we worked out the relative shift 
\be
{\rm  \frac{   Br[\bar{B}\to X_s\gamma \gamma
      ]_{c}^{88} }{ Br[\bar{B}\to X_s\gamma \gamma ] _{c}} } 
\ee 
of the branching ratio due to the (${\cal O}_8$,\,${\cal O}_8$) contribution,
as a function of the kinematical cut parameter $c$.
Fig.~\ref{fig:BR_relativeshift} clearly shows that this contribution is below
$0.1\%$ in the full $(s_1,s_2)$-range considered in this paper. 
We mention that in $\bar{B} \to X_s \gamma$ the situation concerning the 
${\cal O}_8$ contribution is different. As pointed out in refs. 
\cite{Ferroglia:2010xe,Kapustin:1995fk}, in this decay mode the contribution
of ${\cal O}_8$ is non-negligible, in particular, for values of $E_\gamma <
1.1$ GeV. On the other hand, in the double radiative decay, the effects
described in the references just mentioned are also present in the ${\cal O}_7$
contribution; as a consequence the effect of the  ${\cal O}_8$ contribution stays small 
in the full phase space.

\vspace{0.5cm}
\section{Concluding remarks}\label{sec:summary}
In the present work we calculated the set of the $O(\alpha_s)$
corrections to the decay process $\bar{B} \to X_s \gamma \gamma$
originating from diagrams involving the chromomagnetic
dipole operator ${\cal O}_8$.
To perform this calculation, it was necessary to work out diagrams
with four
particles ($s$ quark, two photons and a gluon) in the final state.
From the technical point of view, the calculation was made possible
by the use of the Laporta algorithm \cite{Laporta:2001dd} to identify the
needed master integrals. We then solved the resulting MIs numerically, keeping
the exact dependence on the strange-quark mass $m_{s}$, which we varied
between 400 and 600 $\rm MeV$ in the numerical illustrations.

We conclude that the numerical impact of the self-interference contribution of the chromomagnetic dipole operator ${\cal O}_8$ to the decay 
rate is minor when compared to the self-interference effect of the electromagnetic dipole operator ${\cal O}_7$.

\vspace{2mm}
{\it \bf Acknowledgments} ---
{\small 
H.M.A. acknowledges support from the State Committee of Science of
Armenia Program Grant No.~13-1c153 and Volkswagen Stiftung Program Grant No.~86426.                
C.G. acknowledges the support from the Swiss National Science Foundation.
A.K. acknowledges the support from the United Kingdom Science and Technology Facilities
Council (STFC) under Grant No. ST/L000431/1. 
A.K. thanks Martin Gorbahn for numerous useful discussions.
}

{  \begin{widetext}
\begin{center}
{ \renewcommand{\arraystretch}{1.2}
 \begin{table}[h]
\begin{tabular}{| c | c | c | c || c | c | c |} 
  \hline
  &  \multicolumn{6}{c|}{ Branching ratios for $\bar{B}\to X_s\gamma \gamma$  }                              \\ \hline 
      & \multicolumn{3}{c||}{$c=1/50$}  & \multicolumn{3}{c|}{ $c=1/100$ }     \\       \hline 
&~ $\mu=\,$$m_{b}/2$~~     &~$\mu=$\,$m_{b}$~~   &~$\mu=$\,$2m_{b}$~   &~$\mu=$\,$m_{b}/2$~~    &~$\mu=$\,$m_{b}$~~    &~$\mu=$\,$2m_{b}$~      \\ \hline \hline
   ${\rm NLL_{1}}~$  &1.57        &1.03       &0.71             & 1.79       &1.17       &0.80    \\   \hline
   ${\rm NLL_{2}}$~~&0.96        &0.63       &0.43            &1.09        &0.71       &0.49     \\   \hline
  ${\rm NLL_{3}}$~~&0.59        &0.39         &0.27          &0.67             &0.44       &0.30 
  \\[0.1cm]   \hline \hline
\end{tabular} 
\caption{Branching ratios (in units of $10^{-11}$) for $\bar{B}\to X_s\gamma \gamma$ when only
  considering the (${\cal O}_8$\,,\,${\cal O}_8$) contribution calculated in
  this paper. The left half of the table corresponds to the results when
  choosing $c=1/50$, while in the right half $c$ is set to be $c=1/100$. The
  rows labeled with ${\rm NLL_{1}}$, ${\rm NLL_{2}}$ and ${\rm NLL_{3}}$ give
  the result of this specific NLL contribution when setting $m_{s}=400$ MeV, $m_{s}=500$ MeV and $m_{s}=600$ MeV, respectively. See the text for details.}    
\label{tab:o8o8BRs}
\end{table} 
}        
\end{center}
\end{widetext}      }
%

\section{Appendix}\label{sec:appendix}
In this appendix, we give the explicit formulas defining the four-particle
phase-space region considered in this paper as a result of the intersection
of regions given in \eq{eq:cutsA} and \eq{eq:cuts}. Further, we give the
explicit forms of the master integrals appearing in our calculation of the
$({\cal O}_8,{\cal O}_8)$-contribution to the decay width for $\bar{B} \to X_s
\gamma \gamma$.
\subsection{Explicit formulas for the range in the $(s_1,s_2)$-plane}\label{sec:appendix4Pps}
The kinematical conditions on the phase-space variables $s_1$ and $s_2$, as implicitly
formulated in \eq{eq:cutsA} and \eq{eq:cuts}, can easily be converted to
explicit ranges. There are the following three cases (using $x_4=m_s^2/m_b^2$):
\bea
\label{eq:PScases}
&\nn \underline{ {  \boldmath \large \bf  (i)~{\rm if}~ x_4 \leq c^2} } \\[1mm] \nn
&c<s_{1}<1-2\,c~;~c<s_2<1-s_{1}-c\,\,  \\[2mm] \nn
&{ \underline {  \boldmath \large \bf  (ii)~{\rm if}~c^{2}< x_4 < c \,(1-2\,c)} } \\[1mm]  \nn
&c < s_{1} < \frac { x_4 }{  c } ~;~ \frac { x_4 }{  s_1 } <s_2<1-s_{1}-c\,\,  \\[1mm] 
&~~~{\rm or}  \\[1mm] \nn
& \frac { x_4  }{  c } < s_{1} < 1-2\,c ~;~ c  < s_2<1-s_{1}-c\,\,  \\[2mm] \nn
&{ \underline {  \boldmath \large \bf  (iii)~{\rm if}~x_4  \geq c \,(1-2\,c)} }  \\[1mm] \nn
&{s_1}^{-} <s_{1} < {s_1}^{+} ~\ ; ~\ \frac { x_4 }{  s_1 } < s_2 < 1-s_{1}-c\,\,  \\ \nn
&~~{\rm with}  \\[1mm] \nn
&{s_1}^{\pm}={ \left({1-c \,\pm \, \sqrt{(1-c)^2-4\, x_4  }} \right)/{2}  \, .}
\eea
Case (ii) is understood to be the sum of the two possibilities written in
\eq{eq:PScases}. Further, it can be seen from the same equation 
that if one puts $m_{s}=0$, one would simply end up with case (i), as previously considered in \cite{Asatrian:2011ta,Asatrian:2014mwa}. 
%
\begin{figure}[h]
\begin{center}
\includegraphics[width=4.0cm]{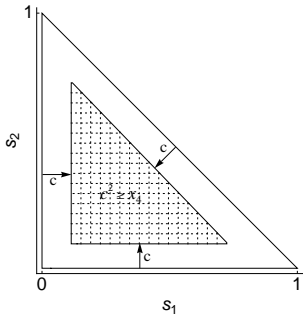}
\caption{The shaded area shows the $(s_1,s_2)$ phase-space region for the case $c^2\geq x_4$.}
\label{fig:phasespace}
\end{center}
\end{figure}
As an example, in Fig.~\ref{fig:phasespace} we give the geometrical representation of case (i) of \eq{eq:PScases}.
\subsection{Explicit forms for the Master Integrals}\label{sec:appendixMIs}
In a first step, we managed to write the triple differential decay width
$d\Gamma_{88}/(ds_1 ds_2 ds_3)$ as a linear combination
of five independent MIs. 

The full four-particle phase space can be parametrized in terms of five independent
variables. According to the procedure described  in Appendix B.2 of
Ref.~\cite{Asatrian:2014mwa}, three of the five variables can be chosen to be
$s_1$, $s_2$ and $s_3$. The MIs are therefore given in terms of
integrals over two variables called $\lambda_4$ and $\lambda_5$, running
in the interval $[0,1]$.

Since we regulated possible collinear singularities by keeping $m_{s}$ exact
and since soft photons are excluded by the cuts imposed through \eq{eq:cuts},
we can work in $d=4$ dimensions; this considerably simplifies the expressions
in Appendix B.2 of Ref.~\cite{Asatrian:2014mwa}.

The MIs, defined at the level of the triple differential decay width, depend on $s_1,s_2,s_3$ and $x_4$.  
We denote them by $B^{\nu_1\nu_2}_{{\rm set}_{i}}(s_1,s_2,s_3,x_4)$, where ${\nu_1,\nu_2}$ stand for the powers of the propagators in the MIs 
and $i$ defines the set (propagator structure) where they belong. Our parametrized MIs are of the form ($\lambda_{4,5} \in [0,1]$)
\bea
 \label{eq:MIs_compact}
 \nn B_{{\rm set}_{i}}^{\nu_1\nu_2}(s_1,s_2,s_3,x_4) =  
 {\cal N}_{\rm ps}  \, \int_{\lambda_4}\, \int_{\lambda_5} d\lambda_{4}d\lambda_{5}   \,  \frac{ P_{1,i}^{-\nu_{1}} P_{2,i}^{-\nu_{2}} }{ \sqrt{ (1-\lambda_{5})\,\lambda_{5} }  }    \,  \\
\eea
where $ {\cal N}_{\rm ps} $ is the phase-space factor with ${\cal N}_{\rm ps}
=  \frac{s_{3}-x_4}{ 2048 \pi^{6} \, s_{3} }$, and the propagators $P_{1,i},
P_{2,i}$ are understood to be expressed in terms of the integration variables
$\lambda_4,\lambda_5$ and the variables $s_1$, $s_2$, $s_3$, following the parametrization used in
\cite{Asatrian:2014mwa}.
Based on these considerations, we have the following expressions for the MIs:
\be
\label{eq:MIs}
\begin{array}{ll}
~~~~~~~~~~~~~~~~~~~~~~~~~~~~~~~\underline{ {\rm set}_{1} }:   \\[1mm]
\underline{   \boldmath P_1=(p_g-p_b+q_1)^2-m_{s}^{2},~P_2=(p_g-p_b)^2-m_{s}^{2} } \\[4mm] 
 B^{00}_{{\rm set}_{1}} \,=\,  \frac{s_{3}-x_4}{ 2048 \pi^{5}  \, s_{3}     }\,\,  \\[2mm] 
 B^{10}_{{\rm set}_{1}} \,=\,   \int_{\lambda_4}\, \int_{\lambda_5} d\lambda_{4}d\lambda_{5}   \,\, \frac{  I^{10}_{ {\rm set}_1}({\lambda_4} ,{\lambda_5} )  }{   \sqrt{\left(1-\lambda _5\right) \lambda _5}  }   \\[2mm]  
\quad \quad  ~=\,    \frac{\log\left(  s_3/x_4  \right)}{ 2048 \pi^{5}\,  (s_{1}-s_{3}) }   \,\,  \\[2mm] 
B^{01}_{{\rm set}_{1}}  \,=\,    \int_{\lambda_4}\, \int_{\lambda_5} d\lambda_{4}d\lambda_{5}   \,\, \frac{  I^{01}_{ {\rm set}_1}({\lambda_4} ,{\lambda_5} )  }{   \sqrt{\left(1-\lambda _5\right) \lambda _5}  }     \,\,  \\[2mm] 
 B^{11}_{{\rm set}_{1}}  \,=\,     \int_{\lambda_4}\, \int_{\lambda_5} d\lambda_{4}d\lambda_{5}   \,\,   \frac{  I^{11}_{ {\rm set}_1}({\lambda_4} ,{\lambda_5} )  }{   \sqrt{\left(1-\lambda _5\right) \lambda _5}  }     \,\,  \\[4mm] 
~~~~~~~~~~~~~~~~~~~~~~~~~~~~~~~\underline{ {\rm set}_{2} }:   \\[1mm] 
\underline{   \boldmath  P_1=(p_{g}-p_b+q_1)^2-m_{s}^{2},~P_2=(p_{g}-p_b+q_{2})^2-m_{s}^{2} } \\[4mm]   
B^{11}_{{\rm set}_{2}}  \,=\,       \int_{\lambda_4}\, \int_{\lambda_5} d\lambda_{4}d\lambda_{5}   \,\,   \frac{  I^{11}_{ {\rm set}_2}({\lambda_4} ,{\lambda_5} )  }{   \sqrt{\left(1-\lambda _5\right) \lambda _5}  }     
\end{array}
\ee
where the respective integrands explicitly read
\bea
\label{eq:integrands10}
&&I^{10}_{ {\rm set}_1} \,= \,  {\cal N}_{\rm ps}    \,  \frac{s_3}{  \left(s_1-s_3\right) \left(s_3 \left(1-\lambda_4\right)+x_4 \lambda _4\right)} \, \, ,  \\[1mm]   \nn
\eea
\bea
   \label{eq:integrands01}
&&\nn  I^{01}_{ {\rm set}_1} \,=  \, {\cal N}_{\rm ps} \,  s_1 \left(s_1-s_3\right) s_3     \, \left[  s_1 \left\{  s_3 \left(   s_1 + \left(s_2-2\right) s_3    \right. \right. \right.   \\ \nn 
&&  \left. \left. \left.   - \left(s_1+s_2-s_3\right) x_4  +  x_4\right)   -    \left(s_1 \left(s_1+s_2\right)  \right. \right. \right. \\  
&& \left.  \left. \left.  -\left(s_1-s_2+2\right) s_3\right) \lambda _4   \left(s_3-x_4\right) \right\}   \right.    \\    \nn
&&\nn \left.   -2  f_{\rm root} \left(s_1-1\right) \left(s_1-s_3\right) \left(2 \lambda _5-1\right)
   \left(s_3-x_4\right)       \right]^{-1}  \, \, , \\[1mm]  \nn
\eea   
\bea
   \label{eq:integrandset111}
   &&  I^{11}_{ {\rm set}_1} \,=\, -    {\cal N}_{\rm ps} \,  s_1 s_3^2  \,  \left[     \left(s_3 \left(\lambda _4-1\right)-x_4 \lambda _4\right)     \right.  \\ 
  &&\nn \left.  \left\{ s_1  \left(s_3
   \left(  s_1 + \left(s_2-2\right) s_3 -\left(s_1+s_2-s_3\right) x_4 + x_4\right)      \right. \right. \right.  \\
  &&\nn \left. \left. \left.  -\left( s_1
   \left(s_1+s_2\right) - \left(s_1-s_2+2\right) s_3\right) \lambda _4 \left(s_3-x_4\right)\right)   \right. \right.  \\
 &&\nn \left. \left.  -2 f_{\rm root} \left(s_1-1\right) \left(s_1-s_3\right) \left(2 \lambda _5-1\right)
   \left(s_3-x_4\right)  \right\}   \right]^{-1}   \, , \\[1mm]  \nn
   \eea
   \bea
    \label{eq:integrandset211}
  && \nn  I^{11}_{ {\rm set}_2} \,=\,  -    {\cal N}_{\rm ps} \,  s_1 s_3^2  \,  \left[      \left(\lambda _4 \left(x_4-s_3\right)+s_3\right) \left\{  2 \left(2 \lambda _5-1\right) \right. \right. \\ \nn
  && \left. \left.  \left(s_1-1\right) \left(s_1-s_3\right) f_{\rm root} \left(s_3-x_4\right)  + s_1  \left(\lambda _4 \left(s_3 \left(s_3-s_2    \right.  \right. \right. \right. \right. \\ \nn
  && \left. \left. \left. \left. \left.  + 2\right)  - s_1 \left(s_2+s_3\right)\right)
   \left(x_4-s_3\right)    + s_3 \left(\left(s_1+s_2-s_3  \right. \right. \right. \right. \right. \\ 
&& \left. \left. \left. \left. \left. -1\right) x_4  -s_1 s_2+s_3\right)\right)  \right\}         \right]^{-1}   \, \, ;   \, 
   \eea
 \be
     \label{eq:root}
   \begin{array}{lllll}
 f_{\rm root} =\,  \sqrt{\frac{s_1^2 \left(s_1+s_2-s_3-1\right) \left(s_1 s_2-s_3\right) s_3 \left(\lambda _4-1\right)
   \lambda _4}{\left(s_1-1\right){}^2 \left(s_1-s_3\right){}^2}}  \,   . \\[1mm]  \nn
   \end{array}
\ee
\vspace{1mm}
In \eq{eq:MIs}, the integrations involved in $ B^{00}_{{\rm set}_{1}}$ were trivial to
perform. For $B^{10}_{{\rm set}_{1}}$, an analytical solution is possible,
using the differential equation (DE) method. 
For the remaining MIs, as the corresponding integrands
$I^{\nu_{1}\nu_{2}}_{ {\rm set}_i}({\lambda_4} ,{\lambda_5})$ develop
complicated structures, we performed these integrations numerically for exact
$m_{s}$.
 
As can be understood from their propagator structures, two of the MIs, $B^{01}_{{\rm set}_{1}}$ and $B^{11}_{{\rm set}_{2}}$, are symmetric under the exchange of $s_{1} \leftrightarrow s_{2}$.

\bibliographystyle{apsrev}
\bibliography{bsgg}

\begin{thebibliography}{46}
\expandafter\ifx\csname natexlab\endcsname\relax\def\natexlab#1{#1}\fi
\expandafter\ifx\csname bibnamefont\endcsname\relax
  \def\bibnamefont#1{#1}\fi
\expandafter\ifx\csname bibfnamefont\endcsname\relax
  \def\bibfnamefont#1{#1}\fi
\expandafter\ifx\csname citenamefont\endcsname\relax
  \def\citenamefont#1{#1}\fi
\expandafter\ifx\csname url\endcsname\relax
  \def\url#1{\texttt{#1}}\fi
\expandafter\ifx\csname urlprefix\endcsname\relax\def\urlprefix{URL }\fi
\providecommand{\bibinfo}[2]{#2}
\providecommand{\eprint}[2][]{\url{#2}}

\bibitem[{\citenamefont{Asatrian et~al.}(2012)\citenamefont{Asatrian, Greub,
  Kokulu, and Yeghiazaryan}}]{Asatrian:2011ta}
\bibinfo{author}{\bibfnamefont{H.}~\bibnamefont{Asatrian}},
  \bibinfo{author}{\bibfnamefont{C.}~\bibnamefont{Greub}},
  \bibinfo{author}{\bibfnamefont{A.}~\bibnamefont{Kokulu}}, \bibnamefont{and}
  \bibinfo{author}{\bibfnamefont{A.}~\bibnamefont{Yeghiazaryan}},
  \bibinfo{journal}{Phys.Rev.} \textbf{\bibinfo{volume}{D85}},
  \bibinfo{pages}{014020} (\bibinfo{year}{2012}), \eprint{1110.1251}.

\bibitem[{\citenamefont{Asatrian and Greub}(2014)}]{Asatrian:2014mwa}
\bibinfo{author}{\bibfnamefont{H.~M.} \bibnamefont{Asatrian}} \bibnamefont{and}
  \bibinfo{author}{\bibfnamefont{C.}~\bibnamefont{Greub}},
  \bibinfo{journal}{Phys.Rev.} \textbf{\bibinfo{volume}{D89}},
  \bibinfo{pages}{094028} (\bibinfo{year}{2014}), \eprint{1403.4502}.

\bibitem[{\citenamefont{Misiak et~al.}(2007)\citenamefont{Misiak, Asatrian,
  Bieri, Czakon, Czarnecki et~al.}}]{Misiak:2006zs}
\bibinfo{author}{\bibfnamefont{M.}~\bibnamefont{Misiak}},
  \bibinfo{author}{\bibfnamefont{H.}~\bibnamefont{Asatrian}},
  \bibinfo{author}{\bibfnamefont{K.}~\bibnamefont{Bieri}},
  \bibinfo{author}{\bibfnamefont{M.}~\bibnamefont{Czakon}},
  \bibinfo{author}{\bibfnamefont{A.}~\bibnamefont{Czarnecki}},
  \bibnamefont{et~al.}, \bibinfo{journal}{Phys.Rev.Lett.}
  \textbf{\bibinfo{volume}{98}}, \bibinfo{pages}{022002}
  (\bibinfo{year}{2007}), \eprint{hep-ph/0609232}.

\bibitem[{\citenamefont{Hurth and Nakao}(2010)}]{Hurth:2010tk}
\bibinfo{author}{\bibfnamefont{T.}~\bibnamefont{Hurth}} \bibnamefont{and}
  \bibinfo{author}{\bibfnamefont{M.}~\bibnamefont{Nakao}},
  \bibinfo{journal}{Ann.Rev.Nucl.Part.Sci.} \textbf{\bibinfo{volume}{60}},
  \bibinfo{pages}{645} (\bibinfo{year}{2010}), \eprint{1005.1224}.

\bibitem[{\citenamefont{Buras}(2011)}]{Buras:2011we}
\bibinfo{author}{\bibfnamefont{A.~J.} \bibnamefont{Buras}}
  (\bibinfo{year}{2011}), \eprint{1102.5650}.

\bibitem[{\citenamefont{Misiak et~al.}(2015)\citenamefont{Misiak, Asatrian,
  Boughezal, Czakon, Ewerth et~al.}}]{Misiak:2015xwa}
\bibinfo{author}{\bibfnamefont{M.}~\bibnamefont{Misiak}},
  \bibinfo{author}{\bibfnamefont{H.}~\bibnamefont{Asatrian}},
  \bibinfo{author}{\bibfnamefont{R.}~\bibnamefont{Boughezal}},
  \bibinfo{author}{\bibfnamefont{M.}~\bibnamefont{Czakon}},
  \bibinfo{author}{\bibfnamefont{T.}~\bibnamefont{Ewerth}},
  \bibnamefont{et~al.}, \bibinfo{journal}{Phys.Rev.Lett.}
  \textbf{\bibinfo{volume}{114}}, \bibinfo{pages}{221801}
  (\bibinfo{year}{2015}), \eprint{1503.01789}.

\bibitem[{\citenamefont{Simma and Wyler}(1990)}]{Simma:1990nr}
\bibinfo{author}{\bibfnamefont{H.}~\bibnamefont{Simma}} \bibnamefont{and}
  \bibinfo{author}{\bibfnamefont{D.}~\bibnamefont{Wyler}},
  \bibinfo{journal}{Nucl.Phys.} \textbf{\bibinfo{volume}{B344}},
  \bibinfo{pages}{283} (\bibinfo{year}{1990}).

\bibitem[{\citenamefont{Reina et~al.}(1997{\natexlab{a}})\citenamefont{Reina,
  Ricciardi, and Soni}}]{Reina:1996up}
\bibinfo{author}{\bibfnamefont{L.}~\bibnamefont{Reina}},
  \bibinfo{author}{\bibfnamefont{G.}~\bibnamefont{Ricciardi}},
  \bibnamefont{and} \bibinfo{author}{\bibfnamefont{A.}~\bibnamefont{Soni}},
  \bibinfo{journal}{Phys.Lett.} \textbf{\bibinfo{volume}{B396}},
  \bibinfo{pages}{231} (\bibinfo{year}{1997}{\natexlab{a}}),
  \eprint{hep-ph/9612387}.

\bibitem[{\citenamefont{Reina et~al.}(1997{\natexlab{b}})\citenamefont{Reina,
  Ricciardi, and Soni}}]{Reina:1997my}
\bibinfo{author}{\bibfnamefont{L.}~\bibnamefont{Reina}},
  \bibinfo{author}{\bibfnamefont{G.}~\bibnamefont{Ricciardi}},
  \bibnamefont{and} \bibinfo{author}{\bibfnamefont{A.}~\bibnamefont{Soni}},
  \bibinfo{journal}{Phys.Rev.} \textbf{\bibinfo{volume}{D56}},
  \bibinfo{pages}{5805} (\bibinfo{year}{1997}{\natexlab{b}}),
  \eprint{hep-ph/9706253}.

\bibitem[{\citenamefont{Cao et~al.}(2001)\citenamefont{Cao, Xiao, and
  Lu}}]{Cao:2001uj}
\bibinfo{author}{\bibfnamefont{J.-j.} \bibnamefont{Cao}},
  \bibinfo{author}{\bibfnamefont{Z.-j.} \bibnamefont{Xiao}}, \bibnamefont{and}
  \bibinfo{author}{\bibfnamefont{G.-r.} \bibnamefont{Lu}},
  \bibinfo{journal}{Phys.Rev.} \textbf{\bibinfo{volume}{D64}},
  \bibinfo{pages}{014012} (\bibinfo{year}{2001}), \eprint{hep-ph/0103154}.

\bibitem[{\citenamefont{Aushev et~al.}(2010)}]{Aushev:2010bq}
\bibinfo{author}{\bibfnamefont{T.}~\bibnamefont{Aushev}} \bibnamefont{et~al.}
  (\bibinfo{year}{2010}), \eprint{1002.5012}.

\bibitem[{sup()}]{superKEKB:online}
\bibinfo{howpublished}{http://www-superkekb.kek.jp/}.

\bibitem[{\citenamefont{Chetyrkin et~al.}(1997)\citenamefont{Chetyrkin, Misiak,
  and Munz}}]{Chetyrkin:1996vx}
\bibinfo{author}{\bibfnamefont{K.~G.} \bibnamefont{Chetyrkin}},
  \bibinfo{author}{\bibfnamefont{M.}~\bibnamefont{Misiak}}, \bibnamefont{and}
  \bibinfo{author}{\bibfnamefont{M.}~\bibnamefont{Munz}},
  \bibinfo{journal}{Phys.Lett.} \textbf{\bibinfo{volume}{B400}},
  \bibinfo{pages}{206} (\bibinfo{year}{1997}), \eprint{hep-ph/9612313}.

\bibitem[{\citenamefont{Kokulu}(2014)}]{Kokulu:2014qqa}
\bibinfo{author}{\bibfnamefont{A.}~\bibnamefont{Kokulu}},
  \bibinfo{journal}{J.Phys.Conf.Ser.} \textbf{\bibinfo{volume}{562}},
  \bibinfo{pages}{012006} (\bibinfo{year}{2014}), \eprint{1411.3763}.

\bibitem[{\citenamefont{Melnikov and Mitov}(2005)}]{Melnikov:2005bx}
\bibinfo{author}{\bibfnamefont{K.}~\bibnamefont{Melnikov}} \bibnamefont{and}
  \bibinfo{author}{\bibfnamefont{A.}~\bibnamefont{Mitov}},
  \bibinfo{journal}{Phys.Lett.} \textbf{\bibinfo{volume}{B620}},
  \bibinfo{pages}{69} (\bibinfo{year}{2005}), \eprint{hep-ph/0505097}.

\bibitem[{\citenamefont{Asatrian et~al.}(2007)\citenamefont{Asatrian, Ewerth,
  Ferroglia, Gambino, and Greub}}]{Asatrian:2006sm}
\bibinfo{author}{\bibfnamefont{H.}~\bibnamefont{Asatrian}},
  \bibinfo{author}{\bibfnamefont{T.}~\bibnamefont{Ewerth}},
  \bibinfo{author}{\bibfnamefont{A.}~\bibnamefont{Ferroglia}},
  \bibinfo{author}{\bibfnamefont{P.}~\bibnamefont{Gambino}}, \bibnamefont{and}
  \bibinfo{author}{\bibfnamefont{C.}~\bibnamefont{Greub}},
  \bibinfo{journal}{Nucl.Phys.} \textbf{\bibinfo{volume}{B762}},
  \bibinfo{pages}{212} (\bibinfo{year}{2007}), \eprint{hep-ph/0607316}.

\bibitem[{\citenamefont{Ferroglia and Haisch}(2010)}]{Ferroglia:2010xe}
\bibinfo{author}{\bibfnamefont{A.}~\bibnamefont{Ferroglia}} \bibnamefont{and}
  \bibinfo{author}{\bibfnamefont{U.}~\bibnamefont{Haisch}},
  \bibinfo{journal}{Phys.Rev.} \textbf{\bibinfo{volume}{D82}},
  \bibinfo{pages}{094012} (\bibinfo{year}{2010}), \eprint{1009.2144}.

\bibitem[{\citenamefont{Misiak and Poradzinski}(2011)}]{Misiak:2010tk}
\bibinfo{author}{\bibfnamefont{M.}~\bibnamefont{Misiak}} \bibnamefont{and}
  \bibinfo{author}{\bibfnamefont{M.}~\bibnamefont{Poradzinski}},
  \bibinfo{journal}{Phys.Rev.} \textbf{\bibinfo{volume}{D83}},
  \bibinfo{pages}{014024} (\bibinfo{year}{2011}), \eprint{1009.5685}.

\bibitem[{\citenamefont{Benzke et~al.}(2010)\citenamefont{Benzke, Lee, Neubert,
  and Paz}}]{Benzke:2010js}
\bibinfo{author}{\bibfnamefont{M.}~\bibnamefont{Benzke}},
  \bibinfo{author}{\bibfnamefont{S.~J.} \bibnamefont{Lee}},
  \bibinfo{author}{\bibfnamefont{M.}~\bibnamefont{Neubert}}, \bibnamefont{and}
  \bibinfo{author}{\bibfnamefont{G.}~\bibnamefont{Paz}},
  \bibinfo{journal}{JHEP} \textbf{\bibinfo{volume}{1008}}, \bibinfo{pages}{099}
  (\bibinfo{year}{2010}), \eprint{1003.5012}.

\bibitem[{\citenamefont{Kapustin et~al.}(1995)\citenamefont{Kapustin, Ligeti,
  and Politzer}}]{Kapustin:1995fk}
\bibinfo{author}{\bibfnamefont{A.}~\bibnamefont{Kapustin}},
  \bibinfo{author}{\bibfnamefont{Z.}~\bibnamefont{Ligeti}}, \bibnamefont{and}
  \bibinfo{author}{\bibfnamefont{H.~D.} \bibnamefont{Politzer}},
  \bibinfo{journal}{Phys.Lett.} \textbf{\bibinfo{volume}{B357}},
  \bibinfo{pages}{653} (\bibinfo{year}{1995}), \eprint{hep-ph/9507248}.

\bibitem[{\citenamefont{Ali and Greub}(1995)}]{Ali:1995bi}
\bibinfo{author}{\bibfnamefont{A.}~\bibnamefont{Ali}} \bibnamefont{and}
  \bibinfo{author}{\bibfnamefont{C.}~\bibnamefont{Greub}},
  \bibinfo{journal}{Phys. Lett.} \textbf{\bibinfo{volume}{B361}},
  \bibinfo{pages}{146} (\bibinfo{year}{1995}), \eprint{hep-ph/9506374}.

\bibitem[{\citenamefont{Kaminski et~al.}(2012)\citenamefont{Kaminski, Misiak,
  and Poradzinski}}]{Kaminski:2012eb}
\bibinfo{author}{\bibfnamefont{M.}~\bibnamefont{Kaminski}},
  \bibinfo{author}{\bibfnamefont{M.}~\bibnamefont{Misiak}}, \bibnamefont{and}
  \bibinfo{author}{\bibfnamefont{M.}~\bibnamefont{Poradzinski}},
  \bibinfo{journal}{Phys. Rev.} \textbf{\bibinfo{volume}{D86}},
  \bibinfo{pages}{094004} (\bibinfo{year}{2012}), \eprint{1209.0965}.

\bibitem[{\citenamefont{Asatrian and Greub}(2013)}]{Asatrian:2013raa}
\bibinfo{author}{\bibfnamefont{H.~M.} \bibnamefont{Asatrian}} \bibnamefont{and}
  \bibinfo{author}{\bibfnamefont{C.}~\bibnamefont{Greub}},
  \bibinfo{journal}{Phys. Rev.} \textbf{\bibinfo{volume}{D88}},
  \bibinfo{pages}{074014} (\bibinfo{year}{2013}), \eprint{1305.6464}.

\bibitem[{\citenamefont{Frixione}(1998)}]{Frixione:1998jh}
\bibinfo{author}{\bibfnamefont{S.}~\bibnamefont{Frixione}},
  \bibinfo{journal}{Phys. Lett.} \textbf{\bibinfo{volume}{B429}},
  \bibinfo{pages}{369} (\bibinfo{year}{1998}), \eprint{hep-ph/9801442}.

\bibitem[{\citenamefont{Gemintern et~al.}(2004)\citenamefont{Gemintern,
  Bar-Shalom, and Eilam}}]{Gemintern:2004bw}
\bibinfo{author}{\bibfnamefont{A.}~\bibnamefont{Gemintern}},
  \bibinfo{author}{\bibfnamefont{S.}~\bibnamefont{Bar-Shalom}},
  \bibnamefont{and} \bibinfo{author}{\bibfnamefont{G.}~\bibnamefont{Eilam}},
  \bibinfo{journal}{Phys.Rev.} \textbf{\bibinfo{volume}{D70}},
  \bibinfo{pages}{035008} (\bibinfo{year}{2004}), \eprint{hep-ph/0404152}.

\bibitem[{\citenamefont{Chang et~al.}(1997)\citenamefont{Chang, Lin, and
  Yao}}]{Chang:1997fs}
\bibinfo{author}{\bibfnamefont{C.-H.~V.} \bibnamefont{Chang}},
  \bibinfo{author}{\bibfnamefont{G.-L.} \bibnamefont{Lin}}, \bibnamefont{and}
  \bibinfo{author}{\bibfnamefont{Y.-P.} \bibnamefont{Yao}},
  \bibinfo{journal}{Phys.Lett.} \textbf{\bibinfo{volume}{B415}},
  \bibinfo{pages}{395} (\bibinfo{year}{1997}), \eprint{hep-ph/9705345}.

\bibitem[{\citenamefont{Hiller and Iltan}(1997)}]{Hiller:1997ie}
\bibinfo{author}{\bibfnamefont{G.}~\bibnamefont{Hiller}} \bibnamefont{and}
  \bibinfo{author}{\bibfnamefont{E.}~\bibnamefont{Iltan}},
  \bibinfo{journal}{Phys.Lett.} \textbf{\bibinfo{volume}{B409}},
  \bibinfo{pages}{425} (\bibinfo{year}{1997}), \eprint{hep-ph/9704385}.

\bibitem[{\citenamefont{Bosch and Buchalla}(2002)}]{Bosch:2002bv}
\bibinfo{author}{\bibfnamefont{S.~W.} \bibnamefont{Bosch}} \bibnamefont{and}
  \bibinfo{author}{\bibfnamefont{G.}~\bibnamefont{Buchalla}},
  \bibinfo{journal}{JHEP} \textbf{\bibinfo{volume}{0208}}, \bibinfo{pages}{054}
  (\bibinfo{year}{2002}), \eprint{hep-ph/0208202}.

\bibitem[{\citenamefont{Bosch}(2002)}]{Bosch:2002bw}
\bibinfo{author}{\bibfnamefont{S.~W.} \bibnamefont{Bosch}}
  (\bibinfo{year}{2002}), \eprint{hep-ph/0208203}.

\bibitem[{\citenamefont{Hiller and Safir}(2005)}]{Hiller:2004wc}
\bibinfo{author}{\bibfnamefont{G.}~\bibnamefont{Hiller}} \bibnamefont{and}
  \bibinfo{author}{\bibfnamefont{A.~S.} \bibnamefont{Safir}},
  \bibinfo{journal}{JHEP} \textbf{\bibinfo{volume}{0502}}, \bibinfo{pages}{011}
  (\bibinfo{year}{2005}), \eprint{hep-ph/0411344}.

\bibitem[{\citenamefont{Hiller and Safir}(2006)}]{Hiller:2005ga}
\bibinfo{author}{\bibfnamefont{G.}~\bibnamefont{Hiller}} \bibnamefont{and}
  \bibinfo{author}{\bibfnamefont{A.~S.} \bibnamefont{Safir}},
  \bibinfo{journal}{PoS} \textbf{\bibinfo{volume}{HEP2005}},
  \bibinfo{pages}{277} (\bibinfo{year}{2006}), \eprint{hep-ph/0511316}.

\bibitem[{\citenamefont{Lin et~al.}(1990)\citenamefont{Lin, Liu, and
  Yao}}]{Lin:1989vj}
\bibinfo{author}{\bibfnamefont{G.-L.} \bibnamefont{Lin}},
  \bibinfo{author}{\bibfnamefont{J.}~\bibnamefont{Liu}}, \bibnamefont{and}
  \bibinfo{author}{\bibfnamefont{Y.-P.} \bibnamefont{Yao}},
  \bibinfo{journal}{Phys.Rev.Lett.} \textbf{\bibinfo{volume}{64}},
  \bibinfo{pages}{1498} (\bibinfo{year}{1990}).

\bibitem[{\citenamefont{Herrlich and Kalinowski}(1992)}]{Herrlich:1991bq}
\bibinfo{author}{\bibfnamefont{S.}~\bibnamefont{Herrlich}} \bibnamefont{and}
  \bibinfo{author}{\bibfnamefont{J.}~\bibnamefont{Kalinowski}},
  \bibinfo{journal}{Nucl.Phys.} \textbf{\bibinfo{volume}{B381}},
  \bibinfo{pages}{501} (\bibinfo{year}{1992}).

\bibitem[{\citenamefont{Choudhury et~al.}(2003)\citenamefont{Choudhury, Joshi,
  Mahajan, and McKellar}}]{Choudhury:2002yu}
\bibinfo{author}{\bibfnamefont{S.}~\bibnamefont{Choudhury}},
  \bibinfo{author}{\bibfnamefont{G.~C.} \bibnamefont{Joshi}},
  \bibinfo{author}{\bibfnamefont{N.}~\bibnamefont{Mahajan}}, \bibnamefont{and}
  \bibinfo{author}{\bibfnamefont{B.}~\bibnamefont{McKellar}},
  \bibinfo{journal}{Phys.Rev.} \textbf{\bibinfo{volume}{D67}},
  \bibinfo{pages}{074016} (\bibinfo{year}{2003}), \eprint{hep-ph/0210160}.

\bibitem[{\citenamefont{Aliev et~al.}(1998)\citenamefont{Aliev, Hiller, and
  Iltan}}]{Aliev:1997uz}
\bibinfo{author}{\bibfnamefont{T.}~\bibnamefont{Aliev}},
  \bibinfo{author}{\bibfnamefont{G.}~\bibnamefont{Hiller}}, \bibnamefont{and}
  \bibinfo{author}{\bibfnamefont{E.}~\bibnamefont{Iltan}},
  \bibinfo{journal}{Nucl.Phys.} \textbf{\bibinfo{volume}{B515}},
  \bibinfo{pages}{321} (\bibinfo{year}{1998}), \eprint{hep-ph/9708382}.

\bibitem[{\citenamefont{Bertolini and Matias}(1998)}]{Bertolini:1998hp}
\bibinfo{author}{\bibfnamefont{S.}~\bibnamefont{Bertolini}} \bibnamefont{and}
  \bibinfo{author}{\bibfnamefont{J.}~\bibnamefont{Matias}},
  \bibinfo{journal}{Phys.Rev.} \textbf{\bibinfo{volume}{D57}},
  \bibinfo{pages}{4197} (\bibinfo{year}{1998}), \eprint{hep-ph/9709330}.

\bibitem[{\citenamefont{Bigi et~al.}(2006)\citenamefont{Bigi, Chiladze,
  Devidze, Hanhart, Lipartelian et~al.}}]{Bigi:2006vc}
\bibinfo{author}{\bibfnamefont{I.~I.} \bibnamefont{Bigi}},
  \bibinfo{author}{\bibfnamefont{G.}~\bibnamefont{Chiladze}},
  \bibinfo{author}{\bibfnamefont{G.}~\bibnamefont{Devidze}},
  \bibinfo{author}{\bibfnamefont{C.}~\bibnamefont{Hanhart}},
  \bibinfo{author}{\bibfnamefont{A.}~\bibnamefont{Lipartelian}},
  \bibnamefont{et~al.}, \bibinfo{journal}{GESJ Phys.}
  \textbf{\bibinfo{volume}{2006N1}}, \bibinfo{pages}{57}
  (\bibinfo{year}{2006}), \eprint{hep-ph/0603160}.

\bibitem[{\citenamefont{Devidze and Jibuti}(1998)}]{Devidze:1998hy}
\bibinfo{author}{\bibfnamefont{G.}~\bibnamefont{Devidze}} \bibnamefont{and}
  \bibinfo{author}{\bibfnamefont{G.}~\bibnamefont{Jibuti}}
  (\bibinfo{year}{1998}), \eprint{hep-ph/9810345}.

\bibitem[{\citenamefont{Aliev and Turan}(1993)}]{Aliev:1993ea}
\bibinfo{author}{\bibfnamefont{T.}~\bibnamefont{Aliev}} \bibnamefont{and}
  \bibinfo{author}{\bibfnamefont{G.}~\bibnamefont{Turan}},
  \bibinfo{journal}{Phys.Rev.} \textbf{\bibinfo{volume}{D48}},
  \bibinfo{pages}{1176} (\bibinfo{year}{1993}).

\bibitem[{\citenamefont{Xiao et~al.}(2003)\citenamefont{Xiao, Lu, and
  Huo}}]{Xiao:2003jn}
\bibinfo{author}{\bibfnamefont{Z.-j.} \bibnamefont{Xiao}},
  \bibinfo{author}{\bibfnamefont{C.-D.} \bibnamefont{Lu}}, \bibnamefont{and}
  \bibinfo{author}{\bibfnamefont{W.-j.} \bibnamefont{Huo}},
  \bibinfo{journal}{Phys.Rev.} \textbf{\bibinfo{volume}{D67}},
  \bibinfo{pages}{094021} (\bibinfo{year}{2003}), \eprint{hep-ph/0301221}.

\bibitem[{\citenamefont{Huo et~al.}(2003)\citenamefont{Huo, Lu, and
  Xiao}}]{Huo:2003cj}
\bibinfo{author}{\bibfnamefont{W.-j.} \bibnamefont{Huo}},
  \bibinfo{author}{\bibfnamefont{C.-D.} \bibnamefont{Lu}}, \bibnamefont{and}
  \bibinfo{author}{\bibfnamefont{Z.-j.} \bibnamefont{Xiao}}
  (\bibinfo{year}{2003}), \eprint{hep-ph/0302177}.

\bibitem[{\citenamefont{Chen and Huo}(2011)}]{Chen:2011te}
\bibinfo{author}{\bibfnamefont{H.}~\bibnamefont{Chen}} \bibnamefont{and}
  \bibinfo{author}{\bibfnamefont{W.}~\bibnamefont{Huo}} (\bibinfo{year}{2011}),
  \eprint{1101.4660}.

\bibitem[{\citenamefont{Qin et~al.}(2009)\citenamefont{Qin, Huo, and
  Yang}}]{Qin:2009zzb}
\bibinfo{author}{\bibfnamefont{X.-M.} \bibnamefont{Qin}},
  \bibinfo{author}{\bibfnamefont{W.-J.} \bibnamefont{Huo}}, \bibnamefont{and}
  \bibinfo{author}{\bibfnamefont{X.-F.} \bibnamefont{Yang}},
  \bibinfo{journal}{Chin. Phys.} \textbf{\bibinfo{volume}{C33}},
  \bibinfo{pages}{252} (\bibinfo{year}{2009}), \eprint{1101.2437}.

\bibitem[{\citenamefont{Ignatiev et~al.}(2005)\citenamefont{Ignatiev, Joshi,
  and McKellar}}]{Ignatiev:2003qm}
\bibinfo{author}{\bibfnamefont{A.~Y.} \bibnamefont{Ignatiev}},
  \bibinfo{author}{\bibfnamefont{G.~C.} \bibnamefont{Joshi}}, \bibnamefont{and}
  \bibinfo{author}{\bibfnamefont{B.}~\bibnamefont{McKellar}},
  \bibinfo{journal}{Int.J.Mod.Phys.} \textbf{\bibinfo{volume}{A20}},
  \bibinfo{pages}{4079} (\bibinfo{year}{2005}), \eprint{hep-ph/0308126}.

\bibitem[{\citenamefont{Asatrian et~al.}(2006)\citenamefont{Asatrian,
  Hovhannisyan, Poghosyan, Ewerth, Greub et~al.}}]{Asatrian:2006ph}
\bibinfo{author}{\bibfnamefont{H.}~\bibnamefont{Asatrian}},
  \bibinfo{author}{\bibfnamefont{A.}~\bibnamefont{Hovhannisyan}},
  \bibinfo{author}{\bibfnamefont{V.}~\bibnamefont{Poghosyan}},
  \bibinfo{author}{\bibfnamefont{T.}~\bibnamefont{Ewerth}},
  \bibinfo{author}{\bibfnamefont{C.}~\bibnamefont{Greub}},
  \bibnamefont{et~al.}, \bibinfo{journal}{Nucl.Phys.}
  \textbf{\bibinfo{volume}{B749}}, \bibinfo{pages}{325} (\bibinfo{year}{2006}),
  \eprint{hep-ph/0605009}.

\bibitem[{\citenamefont{Laporta}(2000)}]{Laporta:2001dd}
\bibinfo{author}{\bibfnamefont{S.}~\bibnamefont{Laporta}},
  \bibinfo{journal}{Int.J.Mod.Phys.} \textbf{\bibinfo{volume}{A15}},
  \bibinfo{pages}{5087} (\bibinfo{year}{2000}), \eprint{hep-ph/0102033}.

\end{thebibliography}

\end{document}